\documentclass[fleqn,twoside,twocolumn,nofootinbib]{revtex4} 
\usepackage[sec]{ujp} 
\begin{document}
\title[OPTICAL POLARIZATION ANISOTROPY, INTRINSIC STARK EFFECT]
{OPTICAL POLARIZATION ANISOTROPY, INTRINSIC\\ STARK EFFECT AND
COULOMB EFFECTS\\ ON THE LASING CHARACTERISTICS\\ OF
 [0001]-ORIENTED GaN/Al\boldmath$_{0.3}$Ga$_{0.7}$N\\ QUANTUM WELLS}%
\author{L.О. Lokot}
\affiliation{V. Lashkaryov Institute of Semiconductor Physics, Nat. Acad. of Sci. of Ukraine}
\address{41, Prosp. Nauky, Kyiv 03028, Ukraine}
\email{llokot@gmail.com}
\udk{533.9} \pacs{73.21.Fg, 77.22.Ej,\\[-3pt] 78.20.H} \razd{\seciv}
\setcounter{page}{12}%
\maketitle

\begin{abstract}
We present a theoretical investigation of space
separated electron and hole distributions, which consists in
the self-consistent solving of the Schr\"{o}dinger equations for electrons and
holes and the Poisson equation. The results are illustrated for
the\linebreak GaN/Al$_{0.3}$Ga$_{0.7}$N quantum well. The optical
gain spectrum in a [0001]-oriented GaN/Al$_{0.3}$Ga$_{0.7}$N quantum
well in the ultraviolet region is calculated. It is found that
both the matrix elements of optical transitions from the heavy hole
band and the optical gain spectrum have only the strict $x$ (or $y$)
light polarization. We present studies of the influence of the confinement of wave
functions on the optical gain which
implicitly depends on the built-in electric field
calculated to be 2.3 MV/cm. Whereas the structures with narrow well
widths exhibit the usual development of the light gain
maximum almost without shifting the spectral region, a significant
blueshift of the gain maximum is found with increase in the plasma
density for wider quantum wells. This blueshift is ascribed to the
interplay between the screening of a strain-induced piezoelectric
field and the bandstructure. A large Sommerfeld or Coulomb enhancement
is present in the quantum well.
\end{abstract}

\section{Introduction}

Direct wide band gap group III-nitride semiconductors based  on GaN
and its alloys have received a great attention due to their
applications in optoelectronic devices such as light-emitting diodes and
lasers at green-blue and near-ultraviolet wavelengths
and solar-blind photodetectors~\cite{{Nakamura},{Savage}}. A number of
ultraviolet light-emitting
diodes~\cite{{Khan2},{Taniyasu},{Mueller},{Wang},{Chu-Kung},{Hirayama}}
and laser diodes~\cite{{Asano},{Lee},{Goddard},{Feltin},{Yoshida}}
already have been demonstrated. Realizing the deep-ultraviolet
semiconductor-based light-emitting diodes will provide compact
high-efficiency light sources for various applications, for example
to the biological detection and data storage ~\cite{{Khan2}}. Thus,
these structures are in the developmental stage.

Here, we present a theoretical investigation of the
intricate interaction of the electron-hole plasma with a built-in
electric field. For this purpose, the calculation of a quantum well bandstructure
is performed using the invariant method and the envelope
approximation. We consider a quantum well of width $w$ in GaN, which
is oriented perpendicularly to the growth direction (0001) and
localized in the spatial region $-w/2<z<w/2$. In the GaN/AlGaN quantum
well structure, there is a strain-induced electric field. This
piezoelectric field, which is perpendicular to the quantum well
plane (i.e., in z direction) may be appreciable because of the large
piezoelectric constants $\hat{e}$ in w\"{u}rtzite structures.

The confinement of wave functions has a strong influence on the optical
gain which is observed with an implicit dependence on the built-in
electric field which is calculated to be 2.3 MV/cm. Such fields are
present in GaN/Al$_{0.3}$Ga$_{0.7}$N systems, because the strain is
induced by the lattice mismatch. The relative magnitude of piezoelectric
effects depends sensitively on the quantum-well width and the plasma
density. In this paper, we present the results of theoretical studies of
the space separation of electron and hole distributions on the basis of
the self-consistent solution of the Schr\"{o}dinger equations for
electrons and holes and the Poisson equation. The Poisson equation
contains the Hartree potential which involves the space
distributions of the charge density for electrons and holes. We focus on details
of the bandstructure for the sake of comparison of
different quantum well structures. In the calculations of a band structure
in the high-concentration regime, we discuss the treatment of the quantum confined Stark effect (QCSE).
By comparing the gain spectra for two GaN/AlGaN quantum well
structures with different well widths, we show the interaction of
the bandstructure and the piezoelectric field. In particular, we will show that
the wide-quantum-well structures, where the QCSE is appreciable, demonstrate a
significant blueshift of the gain maximum, whereas the structures with
a narrow well width exhibit the ordinary behavior of the
light gain maximum almost without shifting the spectral region.

A similar blueshift of the exciton resonance was observed and
analyzed on the microscopic level for GaInN/GaN quantum-well systems
\cite{{Chow}}. They reflect a perturbation of the compensation
between the self-energy and the field renormalization contribution to the
microscopic interband polarization caused by a real spatial distribution of charges.
Such a feature is characteristic of this quantum well and is not inherent
to the GaAs quantum well die to the lacking of a piezoelectric field.
Accounting the Coulomb renormalization of matrix elements of the electric dipole moment in
the two-band model of quantum well structure causes a variation in
the oscillator strength with a variation of the carrier density and the quantum
well configuration.

In work \cite{{Chuang1}}, the matrix
elements of the dipole moment for interband transitions and the optical gain of a deformed w\"{u}rtzite GaN quantum
well were presented without consideration of the intrinsic built-in
piezoelectric field in the quantum well structure.

In work \cite{{Chow1}}, the laser gain was investigated for AlGaN
w\"{u}rtzite quantum well structures. The optical gain spectrum
was computed by simultaneously
diagonalizing the $kp$ Hamiltonian and by solving the Poisson equation. However,
no significant shift of the gain maximum with increase in the plasma
density in the framework of single structure was obtained.
This indicates that, in the given structure in the high-density
regime, QCSE is insignificant. This result coincides with our
calculations of the gain shown in Fig.~5.

In work \cite{{Wang1}}, a self-consistent calculation of the optical
gain in pseudomorphically strained GaN quantum wells
as a function of the carrier density was presented. But
the spectrum renormalization and the electric dipole momentum which are
caused by electron-electron and electron-hole Coulomb correlations were not considered there.

Understanding the influence of the bandstructure and QCSE on laser gain
properties should help one to improve the laser performance and
the optimal configurations of a device.

The light gain spectrum presented in the paper reflects only the
strict TE (\textit{x} or \textit{y}) light polarization. It is known
\cite{{Rashba},{Bir},{Yu},{Sirenko1},{Banal}} that the valence-band
spectrum at the $\Gamma$ point originates from the sixfold
degenerate $\Gamma_{15}$ state. Under the action of the hexagonal
crystal field and the spin-orbit interaction in w\"{u}rtzite
crystals, $\Gamma_{15}$ splits and leads to the formation of three
spin degenerate levels: $\Gamma_{9}$, upper $\Gamma_{7}$, and lower
$\Gamma_{7}$ levels.

In Section 2 for the processes of emission or absorption, we will calculate the
energies  and the wave functions of the lowest conduction
subband and the valence subbands. The
dependences of the matrix elements for
dipole optical interband transitions and the light gain spectrum in
GaN quantum wells on the quantum well width and the charge density are derived. Section 3 presents the Hartree--Fock
light gain spectra and the matrix elements for dipole optical interband
transitions that are calculated within the theory described in Section 2.
By comparing the light gain spectra for two GaN/AlGaN quantum well
structures of different well widths, we show the interaction of
the bandstructure, polarization field, and charge density. We determined
the red renormalization of the light gain spectrum caused by the
electron-electron and hole-hole Coulomb interactions. It is found
that the Sommerfeld enhancement composes 26.7 
gain value, which was obtained in the Hartree problem. This enhancement
of the electric dipole momentum is caused by the electron-hole Coulomb
attraction.

\section{Theory}

We consider QCSE in strained w\"{u}rtzite GaN/Al$_{0.3}$Ga$_{0.7}$N
quantum wells with widths 2.6 nm and 3.9 nm, in which the barrier height
is a constant value for electrons and is equal to $U_{0}=490$
meV. The theoretical analysis of the optical gain of
strained w\"{u}rtzite quantum well lasers is based on
the self-consistent solution of the Schr\"{o}dinger equations for electrons and
holes in quantum well of width $w$ with including Stark effect and the
Poisson equation. The Poisson equation contains the Hartree potential
which involves the charge density for electrons and
holes. All researches are performed at a temperature of 300 K.

The first energy level of an electron in the quantum well of width $w$ is equal to \cite{{Landau}}
\begin{equation}
E_{1}=\frac{2\xi^{2}\hbar^{2}}{mw^{2}},
\end{equation}
where $m=0.19m_{0}$ is an electron effective mass, and $\xi$ is determined
from equation
\begin{equation}
\cos{\xi}=\pm\gamma\xi,
\end{equation}
where $\gamma=\frac{\hbar}{w}\sqrt{\frac{2}{mU_{0}}}$, $\tan{\xi}>0$,
and $\xi=\frac{k_{0}w}{2}$. For $k_{0},$ the following equality holds:
\begin{equation}
\arcsin{\frac{\hbar\,k_{0}}{\sqrt{2mU_{0}}}}=\frac{n\pi-k_{0}w}{2}.
\end{equation}
The wave function of an electron on the first energy level with regard for
QCSE is as follows \cite{{Bastard}}:
\begin{equation}
\Psi(\textbf{r})=\frac{1}{\sqrt{A}}e^{ik_{t}\rho}\Psi(z,\beta)|S\rangle|\sigma_{c}\rangle.
\end{equation}
Here,
\begin{equation}
\Psi(z,\beta)= \left\{ {\begin{array}{l}
 \psi_{1}(z,\beta)\\
\psi(z,\beta)\\
 \psi_{2}(z,\beta)\\
\end{array}} ,\right.
\end{equation}
where
$\psi_{1}(z,\beta)=C_{1}e^{(\kappa_{0}-\beta)(z+\frac{w}{2})}$,
 $\psi(z,\beta)=C\sin{(k_{0}z+\delta_{0})}e^{-\beta\,z}$, $\psi_{2}(z,\beta)=C_{2}e^{-(\kappa_{0}
 +\beta)(z-\frac{w}{2})}$.
From the boundary conditions \cite{{Landau},{Bastard}}
$\psi_{1}(z,\beta)|_{z=-w/2}=\psi(z,\beta)|_{z=-w/2}$,
$\psi_{2}(z,\beta)|_{z=w/2}=\psi(z,\beta)|_{z=w/2}$,
$\frac{\psi'_{1}(z,\beta)}{\psi_{1}(z,\beta)}|_{z=-w/2}=\frac{\psi'(z,\beta)}{\psi(z,\beta)}|_{z=-w/2}$,
$\frac{\psi'_{2}(z,\beta)}{\psi_{2}(z,\beta)}|_{z=w/2}=\frac{\psi'(z,\beta)}{\psi(z,\beta)}|_{z=w/2}$,
we find
$C_{1}=C\sin{(-\frac{k_{0}w}{2}+\delta_{0})}e^{\beta\frac{w}{2}}$,
$C_{2}=C\sin{(\frac{k_{0}w}{2} +\delta_{0})}e^{-\beta\frac{w}{2}}$,
$\kappa_{0}=k_{0}(\frac{1-\cos{k_{0}w}}{\sin{k_{0}w}})$, $\delta_{0}
=\frac{k_{0}w}{2}+\arctan{\frac{\kappa_{0}}{k_{0}}}$, where $A$ is the
area  of a quantum well in the $xy$ plane, $\rho$ is the
two-dimensional vector in the $xy$ plane, and $k_{t}=(k_{x},k_{y})$ is
an in-plane wave vector. The constant multiplier $C$ is found from
the normalization condition
\begin{equation}
\int\limits_{-\infty}^{\infty}|\Psi(z,\beta)|^{2}dz=1.
\end{equation}
Such a representation of the wave function gives the information
that the conduction band corresponds to the $\Gamma_{7}$
representation, which arises due to the splitting of the $C_{6v}^{4}$ space group by the crystal field
with $\Gamma_{1}.$ In other words, the
conduction band wave functions originate from $S$ atomic orbitals.
This is important at the derivation of matrix elements of the electric dipole moment by the
Wigner--Eckart theorem.

The strong mismatch of the lattices in GaN and Al$_{0.3}$Ga$_{0.7}$N leads
to internal strains in the GaN layer. In noncentrosymmetric
structures, the internal strains can induce a macroscopic built-in
polarization field. This phenomenon is known as the piezoelectric
effect. This phenomenon can also be described  as a strain inducing
an electric field. It is known that this piezoelectric field, which
is perpendicular to the quantum well plane, can be significant
because of the large piezoelectric constants in w\"{u}rtzite
structures which are connected with one another:
\begin{equation}
E=-\frac{4\pi}{\kappa}\Big(2\Big(e_{31}-e_{33}\frac{C_{13}}{C_{33}}\Big)\epsilon_{xx}+P_{\rm
sp}\Big),
\end{equation}
where $\hat{e}$ is the piezotensor, $P_{\rm sp}$ is the spontaneous
polarization, $\hat{\epsilon}$ is the strain  tensor, $C_{13}$ and $C_{33}$ are the
elastic constants, and $\kappa$ is the permittivity of the host
material. We calculated the built-in piezoelectric field in the
GaN/Al$_{0.3}$Ga$_{0.7}$N quantum well structure from relation (7) and found
$E\simeq0.23\times 10^{7}$ V/cm.

We take \cite{{Vurgaftman},{Wang1},{Madelung}} the following values
for  constants: $C_{13}=106$ GPa, $C_{33}=398$ GPa,
$e_{31}=-0.44\times 10^{8}$ V/cm, $e_{33}=0.66\times 10^{8}$ V/cm,
$P_{\rm sp}=-0.26\times 10^{7}$ V/cm. The transverse components of
the biaxial strain are proportional to the difference between the
lattice constants of materials of the well and the barrier and depend on the Al
content: $x$, $\epsilon_{xx}=\epsilon_{yy}= \frac{a_{{\rm
Al}_{x}{\rm Ga}_{1-x}{\rm N}}-a_{\rm GaN}}{a_{\rm GaN}}$, $a_{{\rm
Al}_{x}{\rm Ga}_{1-x}{\rm N}}=a_{\rm GaN}+x(a_{\rm AlN}-a_{\rm
GaN})$; $a_{\rm GaN}=0.31892$~nm, $a_{\rm AlN}=0.3112$ nm. The
longitudinal component of a deformation is expressed through elastic
constants and the transverse component of a deformation:
$\epsilon_{zz}=-2\frac{C_{13}}{C_{33}}\epsilon_{xx}$.

One can find the functional, which is built from (4) and (5), in the form
\begin{equation}
J(\beta)=\frac{\langle\Psi|\hat{H}|\Psi\rangle}{\langle\Psi|\Psi\rangle},
\end{equation}
where
\begin{equation}
H=H_{c}-\frac{\hbar^{2}}{2m_{e}^{z}}\frac{\partial^{2}}{\partial\,z^{2}}+V(z),
\end{equation}
$V(z)=U(z)+e\Phi(z)$,
\begin{equation}
H_{c}=E_{g}+\Delta_{1}+\Delta_{2}+\frac{\hbar^{2}}{2m_{e}^{\bot}}k_{t}^{2}+a_{cz}
\epsilon_{zz}+a_{c\bot}(\epsilon_{xx}+\epsilon_{yy}),
\end{equation}
$m_{z,\perp}^{(c)}=0.19 m_{0}$, and $a_{cz,\bot}=-4080~\textrm{meV}$
\cite{Chuang}. The quantity $U(z)$ can be represented in the form
\begin{equation}
U=
\begin{cases}
U_{0}-\frac{eEw}{2}, z\in(-\infty...-w/2),\cr eEz,
z\in[-w/2...w/2],\cr U_{0}+\frac{eEw}{2}, z\in(w/2...\infty).
\end{cases}
\end{equation}
To account the piezoelectric effects, we modify the Schr\"{o}dinger
equation for electrons and holes, by including an off-diagonal
contribution to the electron-hole Hamiltonian. The Schr\"{o}dinger
equation for an infinitely deep quantum well with regard for the QCSE and
the Hartree potential created by spatially separated electrons and
holes can be presented in the form
\begin{equation}
\hat{H}\Psi_{\nu}(\textbf{r})=E_{\nu}\Psi_{\nu}(\textbf{r}),
\end{equation}
where $\hat{H}=\hat{H}_{+}+eEz+e\Phi(z)$. We introduce the Bloch
function written as a vector in the three-dimensional Bloch space:
\begin{equation}
|\alpha\,\sigma_{v}\,k_{t}\rangle=\left\|
\begin{array}{cccc}
\phi_{\alpha}^{(1)}(z,k_{t})\\
\phi_{\alpha}^{(2)}(z,k_{t})\\
\phi_{\alpha}^{(3)}(z,k_{t})\\
\end{array}
\right\|\begin{array}{cccc}
|1,\sigma_{v}\rangle\\
|2,\sigma_{v}\rangle\\
|3,\sigma_{v}\rangle\\
\end{array},\end{equation}
where
\begin{equation}
\phi_{\alpha}^{(j)}=\sum_{i=1}^{n}V_{k_{t}}^{(j)}[i,\alpha]\,\chi_{i}(z),
\end{equation}
and $j=1,2,3$. The Bloch vector of  the $\alpha$-type hole with spin
$\sigma_{v}=\pm$ and momentum $k_{t}$ is specified by its three
coordinates
$[V_{k_{t}}^{(1)}[n,\alpha],\,V_{k_{t}}^{(2)}[n,\alpha],\,V_{k_{t}}^{(3)}[n,\alpha]]$
in the basis
$[|1,\sigma_{v}\rangle,\,|2,\sigma_{v}\rangle,\,|3,\sigma_{v}\rangle]$.
The envelope $z$-dependent part of the quantum well eigenfunctions
can be determined from the boundary conditions
$\chi_{n}(z=-w/2)=\chi_{n}(z=w/2)=0$ for an infinitely deep quantum
well as
\begin{equation}
\chi_{n}(z)=\sqrt{\frac{2}{w}}\,\sin{(\pi\,n\,(\frac{z}{w}+\frac{1}{2}))},
\end{equation}
where $n$ is a natural number.
The hole wave function can be written as
\begin{equation}
\Psi_{\nu}^{v\sigma_{v}}(\textbf{r})=\frac{e^{i\,k_{t}\,\rho}}{\sqrt{S}}\,|\alpha\,\sigma_{v}\,k_{t}\rangle,
\end{equation}
where $\nu=\{k_{t},\alpha\}$ in the envelope-wave approximation,
in which the wave function is considered as a product of
the envelope part $\chi(z)e^{ik_{t}\rho}$ and a periodic Bloch
multiplier. The Bloch vectors in the envelope wave approximation are
projections of the exact Bloch vector on the subspace of vectors with the symmetry inherent to the $\Gamma$
point \cite{Sirenko3}. We have
\begin{equation}
H_{\pm}=\left\|
\begin{array}{cccc}
F & K_{t} & \mp\,iH_{t}\\
K_{t} & G & \Delta\mp\,iH_{t}\\
\pm\,iH_{t} & \Delta\pm\,iH_{t} & \lambda\\
\end{array}
\right\|
\end{equation}
in the basis $[|1,\sigma_{v}\rangle,\,|2,\sigma_{v}\rangle,\,|3,\sigma_{v}\rangle]$ ~\cite{Chang1}, where
\[
F=\Delta_{1}+\Delta_{2}+\lambda+\theta,
\]\vspace*{-5mm}
\[
G=\Delta_{1}-\Delta_{2}+\lambda+\theta,
\]
\[
\lambda=\lambda_{k}+\lambda_{\epsilon},
\]\vspace*{-5mm}
\[
\theta=\theta_{k}+\theta_{\epsilon},
\]\vspace*{-5mm}
\[
\lambda_{k}=\frac{\hbar^{2}}{2m_{0}}(A_{1}k_{z}^{2}+A_{2}k_{t}^{2}),
\]\vspace*{-5mm}
\[
\lambda_{\epsilon}=D_{1}\epsilon_{zz}+D_{2}(\epsilon_{xx}+\epsilon_{yy}),
\]\vspace*{-5mm}
\[
\theta_{k}=\frac{\hbar^{2}}{2m_{0}}(A_{3}k_{z}^{2}+A_{4}k_{t}^{2}),
\]\vspace*{-5mm}
\[
\theta_{\epsilon}=D_{3}\epsilon_{zz}+D_{4}(\epsilon_{xx}+\epsilon_{yy}),
\]\vspace*{-5mm}
\[
K_{t}=\frac{\hbar^{2}}{2m_{0}}(A_{5}k_{t}^{2}),
\]\vspace*{-5mm}
\[
H_{t}=\frac{\hbar^{2}}{2m_{0}}(A_{6}k_{t}k_{z}),
\]\vspace*{-5mm}
\[
\Delta=\sqrt{2}\Delta_{3},
\]\vspace*{-5mm}
\[
k_{t}^{2}=k_{x}^{2}+k_{y}^{2},
\]\vspace*{-5mm}
\[
|1,\pm\rangle=\frac{1}{\sqrt{2}}[|1,1\rangle\,|\uparrow\rangle
\,e^{\frac{-3i\varphi}{2}}e^{-\frac{3i\pi}{4}}\pm|1,-1\rangle\,|\downarrow
\rangle\,e^{\frac{3i\varphi}{2}}e^{\frac{3i\pi}{4}}],
\]\vspace*{-5mm}
\[
|2,\pm\rangle=\frac{1}{\sqrt{2}}[\pm|1,1\rangle\,|\downarrow\rangle
\,e^{\frac{-i\varphi}{2}}e^{-\frac{i\pi}{4}}+|1,-1\rangle\,|\uparrow\rangle
\,e^{\frac{i\varphi}{2}}e^{\frac{i\pi}{4}}],
\]\vspace*{-5mm}
\[
|3,\pm\rangle=\frac{1}{\sqrt{2}}[\pm|1,0\rangle\,|\uparrow\rangle
\,e^{\frac{-i\varphi}{2}}e^{-\frac{i\pi}{4}}+|1,0\rangle\,|\downarrow
\rangle\,e^{\frac{i\varphi}{2}}e^{\frac{i\pi}{4}}],
\]\vspace*{-5mm}
\[
|1,1\rangle=-\frac{1}{\sqrt{2}}|X+iY\rangle,
\]\vspace*{-5mm}
\[
|1,0\rangle=|Z\rangle,\]\vspace*{-5mm}
\[
|1,-1\rangle=\frac{1}{\sqrt{2}}|X-iY\rangle.
\]
The valence subband structure $E_{\alpha}^{\sigma_{v}}(k_{t})$ can
be determined by solving the system of equations
\begin{equation}
\sum_{j=1}^{3}(H_{ij}^{\sigma_{v}}(k_{z}\!=\!-i\,\frac{\partial}{\partial{z}})
\!+\!\delta_{ij}E_{\alpha}^{\sigma_{v}}(k_{t}))\,\phi_{\alpha}^{(j)\sigma_{v}}(z,k_{t})\!=\!0,
\end{equation}
where $i=1,2,3$. In the quasicubic approximation, the parameters of effective mass
and deformation  potential are connected by the relation
~\cite{{Bir},{Sirenko1}}:
\[
4A_{5}-\sqrt{2}A_{6}=A_{3},\quad 2A_{4}=-A_{3}=A_{1}-A_{2},
\]
\[
4D_{5}-\sqrt{2}D_{6}=D_{3},\quad 2D_{4}=-D_{3}=D_{1}-D_{2},
\]
\begin{equation}
\Delta_{2}=\Delta_{3}.
\end{equation}
In calculations, we take the effective-mass parameters for the valence band
\cite{Suzuki} as $A_{1}=-6.56,$ $ A_{2}=-0.91,$ $ A_{3}=5.65,$
$A_{4}=-2.83,$ $ A_{5}=-3.13,$ $ A_{6}=-4.86,$ the parameters for deformation
potential \cite{Chuang} as $D_{1}=700~\textrm{meV},
D_{2}=2100~\textrm{meV},$ $D_{3}=1400~\textrm{meV},
D_{4}=-700~\textrm{meV},$ and the energy parameters at 300~K
\cite{{Vurgaftman},{Chuang1}} as $E_{g}=3507~\textrm{meV},
\Delta_{1}=\Delta_{cr}=16~\textrm{meV},$
$\Delta_{2}=\Delta_{3}=\Delta_{so}/3=4~\textrm{meV}.$ Solving
the Poisson equation
\begin{equation}
\frac{d^{2}\Phi}{dz^{2}}=\frac{4\pi}{\kappa}\rho(z)
\end{equation}
with the condition
$\int_{-\infty}^{\infty}\rho(z)dz=0$ and with the selected wave functions, we find
the Hartree potential $e\Phi(z)$:
\begin{widetext}
\[
e\Phi=\frac{2e^{2}}{\kappa}\sum_{\alpha,n,k,i}g_{\alpha}\int\,k_{t}dk_{t}
\langle\,v_{i},\sigma_{v}|V_{k_{t}}^{i}[\alpha,n]V_{k_{t}}^{i}[\alpha,k]|\sigma_{v},v_{i}\rangle\,f_{\alpha,p}(k_{t})
\begin{cases}
w(\frac{\cos{\pi\,Z(k+n)}}{\pi^{2}(k+n)^{2}}-\frac{\cos{\pi\,Z(n-k)}}{\pi^{2}(n-k)^{2}})\cr
w(\frac{Z^{2}}{2}+\frac{1}{4}\frac{\cos{2\pi\,nZ}}{\pi^{2}n^{2}})
\end{cases}-
\]
\begin{equation}
-\frac{2e^{2}}{\kappa}g_{1}\int\,k_{t}dk_{t}\langle\,S|\langle
\,\sigma_{c}|C^{2}|\sigma_{c}\rangle|S\rangle\,f_{1n}(k_{t})
\begin{cases}
\frac{1-\cos{(-k_{0}w+2\delta_{0})}}{2}e^{\beta\,w}\frac{e^{2(\kappa_{0}-\beta)
(z+\frac{w}{2})}}{4(\kappa_{0}-\beta)^{2}}\cr
\frac{e^{-2\beta\,z}}{8\beta^{2}}-\frac{2\cos{2(k_{0}z+\delta_{0})}e^{-2\beta\,z}}
{(4\beta^{2}+4k_{0}^{2})^{2}}(\beta^{2}-k_{0}^{2})+\frac{\sin{2(k_{0}z+\delta_{0})}e^{-2\beta\,z}}
{4(\beta^{2}+k_{0}^{2})^{2}}k_{0}\beta\cr
\frac{1-\cos{(k_{0}w+2\delta_{0})}}{2}e^{-\beta\,w}\frac{e^{-2(\kappa_{0}+\beta)
(z-\frac{w}{2})}}{4(\kappa_{0}+\beta)^{2}}
\end{cases}\!\!,\end{equation}
\end{widetext}
where $Z=\frac{z}{w}+\frac{1}{2}$, $g_{\alpha}$ and $g_{1}$ correspond to the
degeneration  of the $\alpha$ hole band and
the first quantized conduction band, respectively, $e$ is the value of
electron charge, $\kappa$ is the permittivity of the host material, and
$f_{\alpha,p}(k_{t})$, $f_{1n}(k_{t})$ are the Fermi--Dirac
distributions for holes and electrons. Here we assume the charge
concentrations $9\times 10^{12}~\textrm{cm}^{-2}$, and $7\times
10^{12}~\textrm{cm}^{-2}$.

Solving (12) for holes in the infinitely deep quantum well and finding
the minimum of functional (8) for electrons in a quantum well with
barriers of finite height, we can find the energy and the wave
functions of electrons and holes with regard for the space distribution
of electron and hole charge densities in the quantum well with given
concentrations in a piezoelectric field. The screening field is
determined by iterating Eqs. (8), (12), and (21)
until the convergence of bandstructure calculations is reached. We use the
space carrier distribution of carriers in the lowest order for the envelopes of the wave
functions of electrons and holes.

Consider the matrix elements of interband transitions:
\begin{equation}
M_{j\sigma\rightarrow\,j'\sigma'}(\textbf{k})=\int{d^{3}rU_{j'\sigma'\,\textbf{k}}
\textbf{e}\hat{\mathbf{p}}U_{j\sigma\,\textbf{k}}}.
\end{equation}
The wave functions of the valence band transform according to the
the representation $\Gamma_{1}+\Gamma_{5}$, while the wave function of the conduction band
transforms according to the representation $\Gamma_{1}$. In order to
find the representation for
$M_{j\sigma\rightarrow\,j'\sigma'}(\textbf{k})$, let us consider
the direct product $\Gamma_{1}\times(\Gamma_{1}+\Gamma_{5})$. The symmetry
elements of the point group $C_{6v}$ are as follows:
\begin{equation}
g=E, C_{2}, 2C_{3}, 2C_{6}, 3\sigma_{v}, 3\sigma_{v}',
\end{equation}
where $C_{n}$ is the axis of the $n$-th order, $3\sigma_{v}$ and $3\sigma_{v}'$ are 6 planes
of reflection which pass through the sixth-order axis. For these
elements, we find the representation $\Gamma_{1}+\Gamma_{5}$:
\begin{equation}
\begin{array}{l}
\chi(E)=3, \chi(C_{2})=-1, \chi(2C_{3})=0,\\ [1.5mm]
\chi(2C_{6})=2, \chi(3\sigma_{v})=1, \chi(3\sigma_{v}')=1.\\
\end{array}
\end{equation}
The squares of irreducible representation elements are
\begin{equation}
g^{2}=E, E, C_{3}, C_{3}, E, E.
\end{equation}
We need to find
\begin{equation}
\begin{array}{l}
\chi_{\psi}^{2}(E)=9, \chi_{\psi}^{2}(C_{2})=1,
\chi_{\psi}^{2}(2C_{3})=0,\\ [1.5mm]
\chi_{\psi}^{2}(2C_{6})=4, \chi_{\psi}^{2}(3\sigma_{v})=1, \chi_{\psi}^{2}(3\sigma_{v}')=1,\\
\end{array}
\end{equation}
whereas
\begin{equation}
\begin{array}{l}
\chi_{\psi}(E^{2})=3, \chi_{\psi}(C_{2}^{2})=3,
\chi_{\psi}(2C_{3}^{2})=0,\\ [1.5mm]
\chi_{\psi}(2C_{6}^{2})=0, \chi_{\psi}(3\sigma_{v}^{2})=3, \chi_{\psi}(3\sigma_{v}'^{2})=3.\\
\end{array}
\end{equation}
The symmetric representation can be found in the form
\begin{equation}
\begin{array}{l}
\frac{1}{2}(\chi_{\psi}^{2}(g)+\chi_{\psi}(g^{2})):\\ [1.5mm]
\frac{1}{2}(\chi_{\psi}^{2}(E)+\chi_{\psi}(E^{2}))=6,\\ [1.5mm]
\frac{1}{2}(\chi_{\psi}^{2}(C_{2})+\chi_{\psi}(C_{2}^{2}))=2,\\
[1.5mm]
\frac{1}{2}(\chi_{\psi}^{2}(2C_{3})+\chi_{\psi}(2C_{3}^{2}))=0,\\
[1.5mm]
\frac{1}{2}(\chi_{\psi}^{2}(2C_{6})+\chi_{\psi}(2C_{6}^{2}))=2,\\
[1.5mm]
\frac{1}{2}(\chi_{\psi}^{2}(3\sigma_{v})+\chi_{\psi}(3\sigma_{v}^{2}))=2,\\
[1.5mm]
\frac{1}{2}(\chi_{\psi}^{2}(3\sigma_{v}')+\chi_{\psi}(3\sigma_{v}'^{2}))=2.\\
\end{array}
\end{equation}
The antisymmetric representations are
\begin{equation}
\begin{array}{l}
\frac{1}{2}(\chi_{\psi}^{2}(g)-\chi_{\psi}(g^{2})):\\ [1.5mm]
\frac{1}{2}(\chi_{\psi}^{2}(E)-\chi_{\psi}(E^{2}))=3,\\ [1.5mm]
\frac{1}{2}(\chi_{\psi}^{2}(C_{2})-\chi_{\psi}(C_{2}^{2}))=-1,\\
[1.5mm]
\frac{1}{2}(\chi_{\psi}^{2}(2C_{3})-\chi_{\psi}(2C_{3}^{2}))=0,\\
[1.5mm]
\frac{1}{2}(\chi_{\psi}^{2}(2C_{6})-\chi_{\psi}(2C_{6}^{2}))=2,\\
[1.5mm]
\frac{1}{2}(\chi_{\psi}^{2}(3\sigma_{v})-\chi_{\psi}(3\sigma_{v}^{2}))=-1,\\
[1.5mm]
\frac{1}{2}(\chi_{\psi}^{2}(3\sigma_{v}')-\chi_{\psi}(3\sigma_{v}'^{2}))=-1.\\
\end{array}
\end{equation}
The symmetric representation can be decomposed into the
irreducible representations $2A_{1}+E_{1}+E_{2}$, whereas the
antisymmetric one into $A_{2}+E_{1}$. Thus, the w\"{u}rtzite
Hamiltonian $H(\epsilon,\textbf{k})$ must include the even
functions (with respect to the time inversion), which are
transformed according to $2A_{1}+E_{1}+E_{2},$ and odd functions,
which are transformed according to $A_{2}+E_{1}$ ~\cite{Bir}.

The vector representation can be written as
\begin{equation}
\begin{array}{l}
\chi_{v}(E)=3, \chi_{v}(C_{2})=-1, \chi_{v}(2C_{3})=0,\\ [1.5mm]
\chi_{v}(2C_{6})=2, \chi_{v}(3\sigma_{v})=1, \chi_{v}(3\sigma_{v}')=1,\\
\end{array}
\end{equation}
and can be decomposed into the irreducible representations $A_{1}+E_{1}$.
The representation, according to which the interband operator is transformed, can be decomposed into
\begin{equation}
\Gamma_{1}\times(\Gamma_{1}+\Gamma_{5})=A_{1}+E_{1}.
\end{equation}

Thus, the direct product of representations (31) reflects the existence of
nonzero  matrix elements of the electric dipole moment of interband
transitions, because the
vector representation can be formed from these representations.\looseness=1

Allowed matrix elements of the electric dipole moment $\langle\,S|\langle\,\sigma_{c}|
\textbf{e}\hat{\mathbf{p}}|v_{i},\sigma_{v}\rangle$ are found in the form
\begin{equation}
\begin{array}{l}
\langle\,S|\langle\,\uparrow|\textbf{e}\hat{\mathbf{p}}|v_{1},\pm\rangle
=-\frac{1}{2}P_{\bot}e^{i\varphi}e^{-i\frac{3\pi}{4}}\sin{\theta},\\
[1.5mm]
\langle\,S|\langle\,\uparrow|\textbf{e}\hat{\mathbf{p}}|v_{2},\pm\rangle
=\frac{1}{2}P_{\bot}e^{-i\varphi}e^{i\frac{\pi}{4}}\sin{\theta},\\
[1.5mm]
\langle\,S|\langle\,\uparrow|\textbf{e}\hat{\mathbf{p}}|v_{3},\pm\rangle
=\pm\frac{1}{\sqrt{2}}P_{z}e^{-i\frac{\pi}{4}}\cos{\theta},\\
[1.5mm]
\langle\,S|\langle\,\downarrow\,|\textbf{e}\hat{\mathbf{p}}|v_{1},\pm\rangle
=\pm\frac{1}{2}P_{\bot}e^{-i\varphi}e^{i\frac{3\pi}{4}}\sin{\theta},\\
[1.5mm]
\langle\,S|\langle\,\downarrow|\textbf{e}\hat{\mathbf{p}}|v_{2},\pm\rangle
=\mp\frac{1}{2}P_{\bot}e^{i\varphi}e^{-i\frac{\pi}{4}}\sin{\theta},\\
[1.5mm]
\langle\,S|\langle\,\downarrow|\textbf{e}\hat{\mathbf{p}}|v_{3},\pm\rangle
=\frac{1}{\sqrt{2}}P_{z}e^{i\frac{\pi}{4}}\cos{\theta}.\\
\end{array}
\end{equation}
Due to the symmetry properties of the Bloch functions, the only nonzero matrix elements between the basis functions are \cite{{Landau},{Sirenko1}}
\begin{equation}
\begin{array}{l}
\langle\,S|\hat{p}_{z}|1,0\rangle=P_{z},\\ [1.5mm]
\langle\,S|\hat{p}_{+}|1,-1\rangle=-\langle\,S|\hat{p}_{-}|1,1\rangle=\sqrt{2}\,P_{\perp},\\
\end{array}
\end{equation}
where
$\hat{p}_{\pm}=\hat{p}_{x}\pm{i}\, \hat{p}_{y}$. Two constants of the matrix
elements of the moment are as follows:
$P_{\perp}\equiv\langle\,S|\hat{p}_{x}|X\rangle$ and
$P_{z}\equiv\langle\,S|\hat{p}_{z}|Z\rangle$. Due to the cylindrical
symmetry, the matrix element depends only on the difference
$\varphi=\varphi_{\textbf{E}}-\varphi_{\textbf{k}}$ between
the plane-projected angles of the vectors $\textbf{e}\|\textbf{E}$ and
$\textbf{k}$. To simplify calculations, we assume
$\varphi_{\textbf{k}}=0$ and denote the spherical angles of the vector
$\textbf{e}$ by $\varphi$ and $\theta$ \cite{{Sirenko1}}. We
consider the case of a hole wave vector parallel to the $c$ axis. In
this situation, $\varphi=0$ in our calculations, and the vector
$\textbf{e}$ in the spherical coordinates takes the form
$\textbf{e}=(\sin{\theta}\cos{\varphi},\sin{\theta}\sin{\varphi},\cos{\theta})$,
whereas
$\mathbf{e}\hat{\mathbf{p}}=\frac{1}{2}\sin{\theta}(e^{i\varphi}\hat{p}_{-}+e^{-i\varphi}\hat{p}_{+})+\hat{p}_{z}\cos{\theta}$.
It is known~\cite{Sirenko1} that the values of constants
$|P_{z,\perp}|^{2}$ can be found from the \textit{kp} theory:
\begin{equation}
\frac{m_{0}}{m_{z,x}^{(c)}}=1+\frac{2}{m_{0}}\Sigma_{j\neq{c}}\frac{|\langle{c}|
\hat{p}_{z,x}|j\rangle|^2}{E_{c}^{0}-E_{j}^{0}}.
\end{equation}
From the value of experimentally measured conduction-band effective mass
$m_{z,\perp}^{(c)}=0.19 m_{0}$ and $E_{g}=3.5$ eV, we obtain
$\frac{2|P_{z,\bot}|^{2}}{m_{0}}\sim 15$ eV.

\begin{figure}
\includegraphics[width=\column]{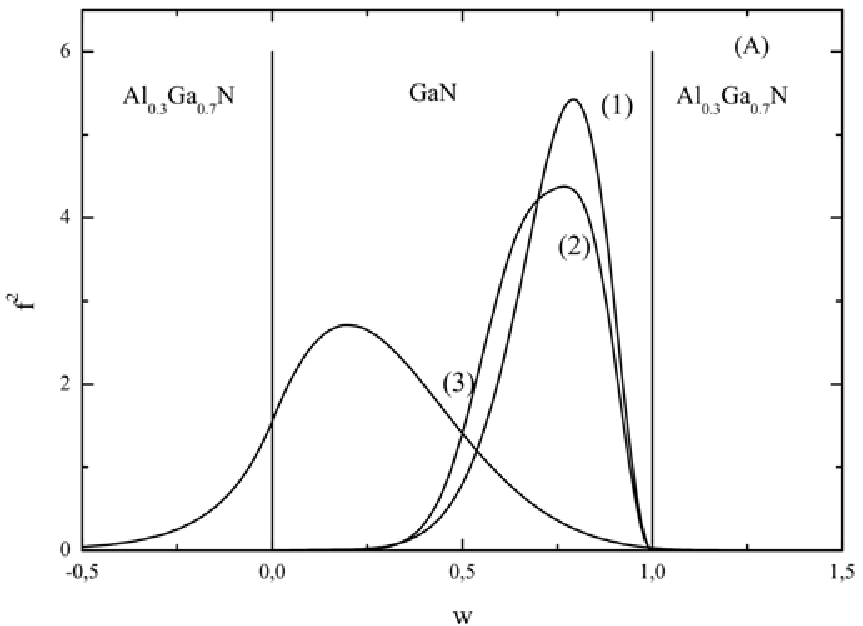}\\[2mm]
\includegraphics[width=\column]{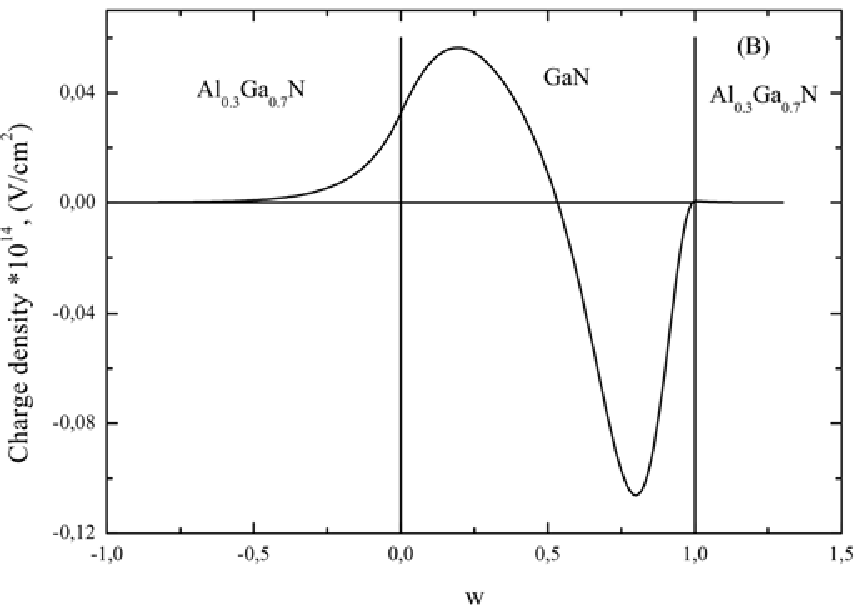}
\vskip-3mm\caption{Calculated square of the wave functions of a
heavy hole ({\it 1}) and a light hole ({\it 2}) at the transverse
wave vector $k_{t}=8\times 10^{6} \textrm{cm}^{-1}$ and an
electron ({\it 3}) (A); the charge density distribution on the
quantum well 3.9 nm in width (B) at the charge concentration
$9\times 10^{12}\textrm{cm}^{-2}$}
\end{figure}
\begin{figure}
\includegraphics[width=\column]{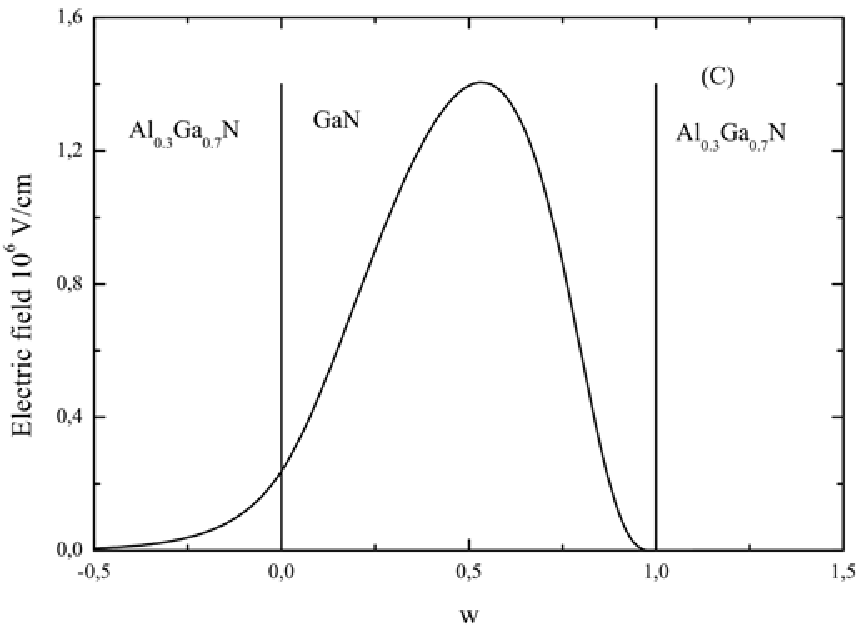}\\[2mm]
\includegraphics[width=\column]{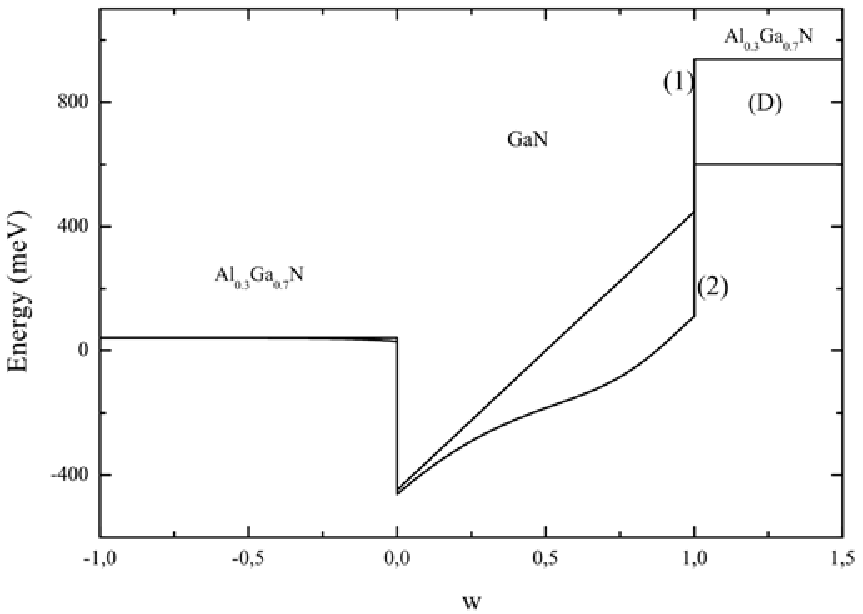}
\vskip-3mm\caption{Effective screening electric field distribution
(C); the quantum well potential (\textit{1}) and screening
potential (\textit{2}) on the quantum well 3.9 nm in width (D) at
the charge concentration $9\times 10^{12}~\textrm{cm}^{-2}$}
\end{figure}

In Fig. 3, we show the $k$-dependence of the matrix elements for the
quantum well.  We see that the matrix elements have the strict $x$
(or $y$) light polarization for the transitions from the heavy hole
band to the conduction band, while for the $z$ light polarization, these transitions are
forbidden \cite{{Lokot}}. That is why the light gain spectra
presented in Figs. 4 and 5 reflect only the gain of TE polarized light
for two widths of the quantum well. Such a behavior agrees with the results of
calculations of the moment matrix elements for a w\"{u}rtzite GaN
quantum well, since the valence band top originates from
$\Gamma_{9}$, $\Gamma_{7}$, and $\Gamma_{7}$ irreducible
representations. The results which are presented in Figs. 3--5,
testify to the optical polarization anisotropy of the matrix
elements of the electric dipole moment for interband transitions in GaN/Al$_{0.3}$Ga$_{0.7}$N
quantum well structures.

\begin{figure}
\includegraphics[width=\column]{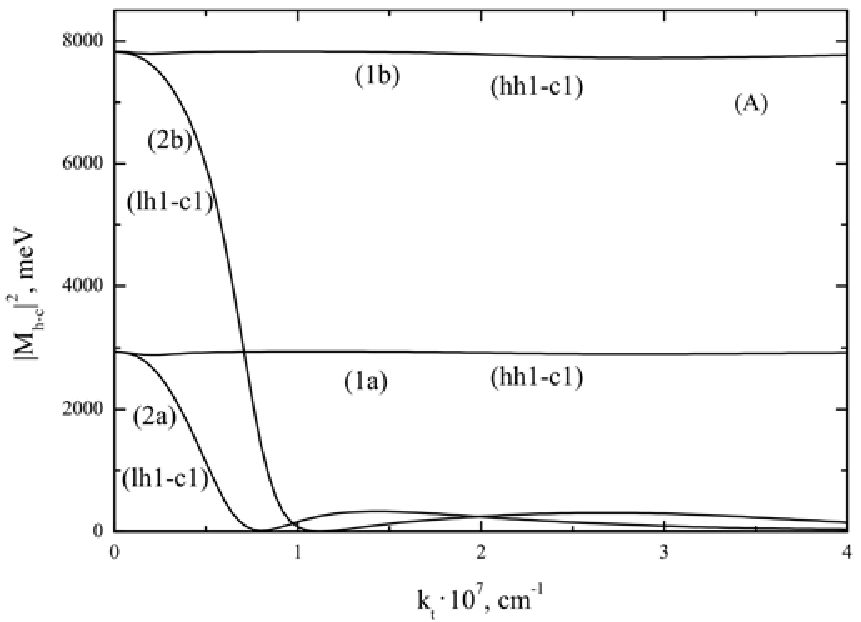}\\[2mm]
\includegraphics[width=\column]{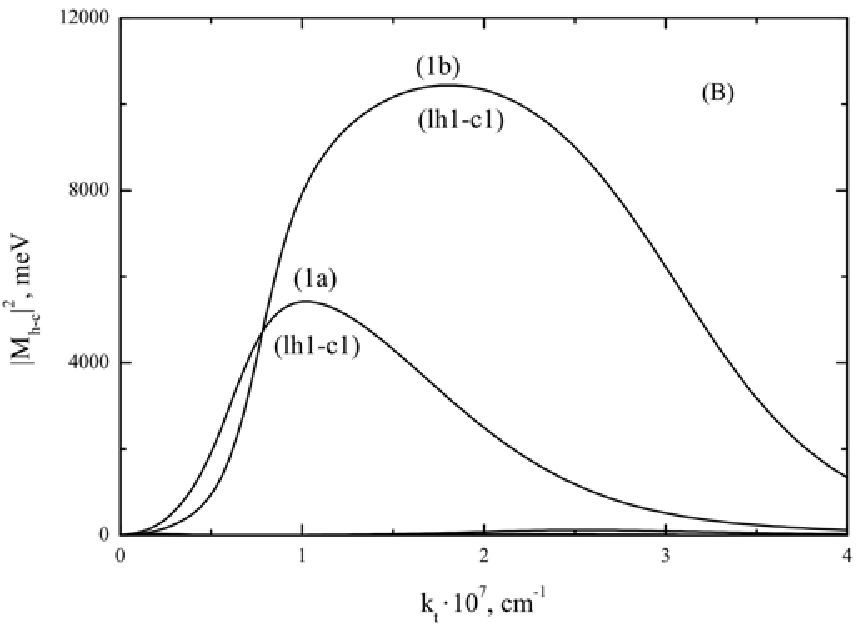}
\vskip-3mm\caption{Moment matrix elements for the \textit{x}-
(or \textit{y}-) polarization (A) and the $z$-polarization (B):
(1\textit{a}) quantum well 3.9~nm in width  at the concentration
$n=p=9\times 10^{12}~\textrm{cm}^{-2}$; (2\textit{a}) quantum well
of width 3.9 nm at the concentration $n=p=7\times
10^{12}~\textrm{cm}^{-2}$; (1\textit{b}) quantum well
2.6~nm in width at the concentration $n=p=9\times 10^{12}~\textrm{cm}^{-2}$;
(2{\it b}) quantum well 2.6~nm in width at the concentration
$n=p=7\times 10^{12}~\textrm{cm}^{-2}$}
\end{figure}

The optical gain of a material \cite{{Kochelap},{Chuang1}} can be
calculated from the Fermi golden rule
\[
\alpha_{0}=\frac{\pi\,e^{2}}{c\,\sqrt{\kappa}\,m_{0}\,w\,\omega}\times
\]
\[
\times\sum_{\sigma_{c}
=\uparrow,\downarrow}\sum_{\sigma_{v}=+,-}\sum_{m,\alpha}\int{k_{t}\,dk_{t}}\int{\frac{d\phi}{2\,\pi}}\,|\hat{\mathbf{e}}\,
M_{m\,\alpha}^{\sigma_{c}\,\sigma_{v}}(k_{t})|^{2}\times
\]
\begin{equation}
\times\,\frac{(f_{m}^{c}(k_{t})-f_{\sigma_{v}\alpha}^{v}(k_{t}))(\frac{\hbar\,\gamma}{\pi})}
{(E_{\sigma_{v},m\alpha}^{cv}(k_{t})-\hbar\,\omega)^{2}+(\hbar\,\gamma)^{2}},
\end{equation}
where $e$ is the magnitude of electron charge, $m_{0}$ is the
electron rest mass in the free space, $c$ is the velocity of light in
the free space, $\kappa=8.27$ is the permittivity of the host material,
$f_{m}^{c}$ and $f_{\sigma_{v}\alpha}^{v}$ are the Fermi--Dirac
distributions for electrons in the conduction and valence bands, respectively,
$\textbf{e}$ is a unit vector of the vector potential of the electromagnetic
field, $E_{\sigma_{v},m\alpha}^{cv}(k_{t})$ is the interband energy
of the conduction and valence bands, $\hbar\,\omega$ is the
optical energy, and $\hbar\,\gamma$ is a half-linewidth of the
Lorentzian function, which is equal 6.56 meV. We consider the
electromagnetic wave which propagates in the plane of the quantum well. The
modal gain, which determines the threshold condition of a laser, is
proportional to the material gain multiplied by the optical
confinement factor $\Gamma$ and by the number $a$ of quantum wells
in the case of multiple quantum wells: $\alpha=\alpha_{0}\,\Gamma\,a$. We
take $\Gamma$ equal to 0.01, and $a$ is taken to be 1 in the
calculations.\looseness=1

\begin{figure}
\includegraphics*[width=\column]{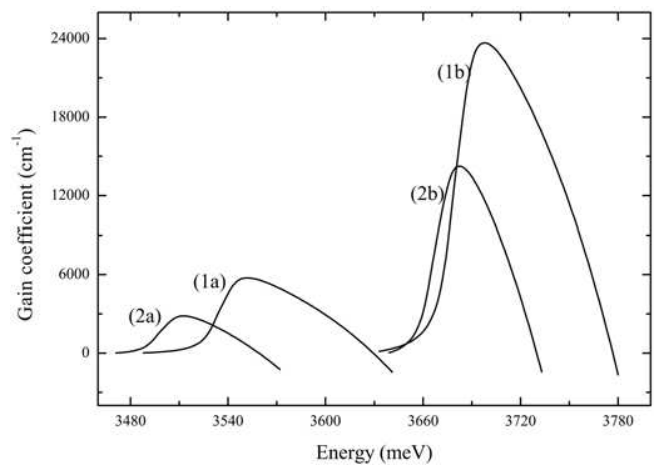}
\vskip-3mm\caption{Calculated Hartree gain spectrum: (1a) quantum
well 3.9~nm in width at the concentration $n=p=9\times
10^{12}~\textrm{cm}^{-2}$; (2\textit{a}) quantum well
3.9~nm in width at the concentration $n=p=7\times 10^{12}\textrm{cm}^{-2}$;
(1\textit{b}) quantum well 2.6~nm in width at the concentration
$n=p=9\times 10^{12}~\textrm{cm}^{-2}$; (2\textit{b}) quantum well
2.6 nm in width at the concentration $n=p=7\times
10^{12}~\textrm{cm}^{-2}$}\vskip3mm
\end{figure}
\begin{figure}
\includegraphics[width=\column]{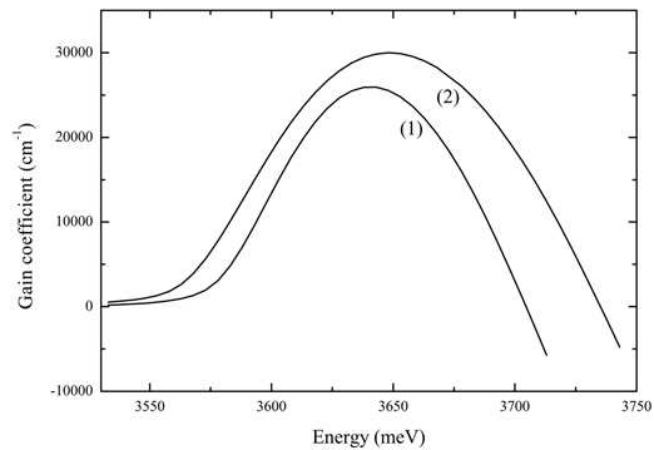}
\vskip-3mm\caption{Calculated Hartree--Fock gain spectrum for the
quantum well 2.6 nm in width and at concentrations: $n=p=7\times
10^{12}~\textrm{cm}^{-2}$ ({\it 1}) and $n=p=9\times
10^{12}~\textrm{cm}^{-2}$ ({\it 2})}
\end{figure}

Although the carriers within each band are in a strongly
nonequilibrium state,  the interband relaxation times are much
larger than intraband relaxation times. Therefore, the Fermi--Dirac
statistics can be used in the calculations.

Using the expressions for the basis functions, we obtain two scalar
polarizations for the matrix elements of the electric dipole
moment. For the TE-polarization ($\hat{e}=\hat{x}$ or $\hat{y} \bot$ $c$
axis), i.e., for the light polarization vector lying in the
quantum well plane, we have
\begin{equation}
\begin{array}{l}
|(M_{x})_{1\alpha}^{\sigma}(k_{t})|^{2}=\\[1.5mm]=\frac{|\langle\,S|p_{x}|X\rangle|^{2}}{4}\{\langle\Psi_{1}(\beta)|\sum_{n}V_{k_{t}}^{1}
[n,\alpha]\chi_{n}\rangle^{2}+\\ [1.5mm]
+\langle\Psi_{1}(\beta)|\sum_{n}V_{k_{t}}^{2}[n,\alpha]\chi_{n}\rangle^{2}\},\\
[1.5mm] \textrm{for}\, \sigma=+,\\
[1.5mm]=\frac{|\langle\,S|p_{x}|X\rangle|^{2}}{4}\{\langle\Psi_{1}(\beta)|\sum_{n}V_{k_{t}}^{4}[n,\alpha]
\chi_{n}\rangle^{2}+\\ [1.5mm]
+\langle\Psi_{1}(\beta)|\sum_{n}V_{k_{t}}^{5}[n,\alpha]\chi_{n}\rangle^{2}\},\\
[1.5mm]
\textrm{for}\, \sigma=-.\\
\end{array}
\end{equation}
For the TM-polarization ($\hat{e}=\hat{z}\|$$c$ axis), i.e., for the light
polarization vector, which is perpendicular to the quantum well
plane, we have
\begin{equation}
\begin{array}{l}
|(M_{z})_{1\alpha}^{\sigma}(k_{t})|^{2}=\\ [1.5mm]
=\frac{|\langle\,S|p_{z}|Z\rangle|^{2}}{2}\{\langle\Psi_{1}(\beta)|\sum_{n}V_{k_{t}}^{3}
[n,\alpha]\chi_{n}\rangle^{2}\},\\ [1.5mm] \textrm{for}\,
\sigma=+,\\
[1.5mm]=\frac{|\langle\,S|p_{z}|Z\rangle|^{2}}{2}\{\langle\Psi_{1}(\beta)|\sum_{n}V_{k_{t}}^{6}
[n,\alpha]\chi_{n}\rangle^{2}\},\\ [1.5mm]
\textrm{for}\, \sigma=-.\\
\end{array}
\end{equation}

\section{Results and Their Discussion}

To describe the interplay of the bandstructure and the polarization
effects in the Hartree problem, we consider the 2.6-nm and 3.9-nm
GaN/Al$_{0.3}$Ga$_{0.7}$N quantum well structures. In the quantum
well 2.6 nm in width at a concentration of $9\times
10^{12}~\textrm{cm}^{-2}$, a optical gain maximum is equal to
$23673.7 ~\textrm{cm}^{-1}$ at the wavelength  $\lambda=334.5$ nm;
while, at a concentration of $7\times 10^{12}~\textrm{cm}^{-2}$, the
optical gain maximum is equal to $14245.9 ~\textrm{cm}^{-1}$. Such a
gain is observed at the wavelength  $\lambda=336$ nm. In the quantum
well 3.9 nm in width at a concentration of $9\times
10^{12}~\textrm{cm}^{-2}$, the optical gain maximum is equal to
$5752.7 \textrm{cm}^{-1}$ at the wavelength  $\lambda=348.3$ nm;
while, at a concentration of $7\times 10^{12}~\textrm{cm}^{-2}$, the
optical gain maximum is equal to $2840.8 ~\textrm{cm}^{-1}$. Such a
gain is calculated at the wavelength  $\lambda=352.1$ nm. Thus, the
optical gain in the GaN/Al$_{0.3}$Ga$_{0.7}$N quantum well develops
in the ultraviolet spectral region, as shown in Fig. 4.

Numerically solving the Schr\"{o}dinger equations (8) and (12) for
electrons and holes and the Poisson equation (20), the steady state
solutions allow us to construct the squares of the wave functions of
a heavy hole (\textit{1}), a light hole (\textit{2}) (e.g., at the
transverse wave vector $k_{t}=8\times 10^{6}~ \textrm{cm}^{-1}$),
and an electron (\textit{3}) (A); the charge density distribution on
the quantum well width (B); the effective screening electric field
distribution (C); and the quantum well potential and the screening
potential on the quantum well width (D). The results of calculations
for the concentration $9\times 10^{12}~\textrm{cm}^{-2}$ for a
3.9-nm quantum well are shown in Figs. 1 and 2, and the Hartree gain
spectra are presented in Fig. 4. For the narrow quantum well, we see
that, at a density of $7\times 10^{12}\textrm{cm}^{-2},$ the light
gain is gradually developed, as the carrier density increases. At
high densities (i.e., when the density is equal to $9\times
10^{12}~\textrm{cm}^{-2}$), the optical gain develops nearly in the
spectral region of the original optical gain at a plasma density of
$7\times 10^{12}~\textrm{cm}^{-2}$.

The behavior of the light gain coefficient for two quantum well widths and
at given concentrations can be understood from Figs. 1(A,B) and 2(C,D).
From Fig. 1(A), we see that the overlap between the quantum confined
electron and hole wave functions is related to the charge density
distribution over the quantum well width, which is shown in Fig.
1(B). We can conclude that, for the wide quantum well, the overlapping integral is smaller than that for a
narrow quantum well and reduces stronger with decrease in the carrier
density.

The effective screening electric field distribution for
wide-bandgap GaN/AlGaN quantum well systems, which is presented in Fig.
2(C), is similar to that of the electric field in a condenser.

As shown in Fig. 4, the situation is quite different for the 3.9-nm
GaN/Al$_{0.3}$Ga$_{0.7}$N quantum well structure. Because of a
weaker quantum confinement in this relatively wide quantum well, the
piezoelectric field is able to significantly reduce the overlap
between the quantum confined electron and hole wave functions, which
can be seen from the comparison of Figs. 1 and 2. As a result the interband dipole matrix
element or the oscillator strength is substantially smaller than that in the
case for the narrow 2.6-nm quantum well. This intrinsic quantum
confined Stark effect can also cause a significant redshift of the gain maximum
at a plasma density of $7\times 10^{12}~\textrm{cm}^{-2}$ as compare with the
flat-bottom band situation. As the plasma density increases, the screening
of the QCSE increases the overlap of electron-hole wave functions and,
hence, the exciton oscillator strength. Simultaneously,
the weakened piezoelectric field, which induced earlier the redshift, leads to
the net of blueshifts in the gain maximum and the absorption edge with
increase in the plasma density, as shown in Fig. 4.

To calculate the concentration dependence of many-body Coulomb
effects in the absorption  spectrum of a GaN quantum well, we apply
the method developed in \cite{Lindberg,Chow,Chowbook}. Numerically
solving the microscopic polarization equation, we see that, with
increase in the plasma density, the optical gain (i.e., negative
absorption) develops in the spectral region of the original exciton
resonance. With increase in the free-carrier density, the ionization
continuum shifts rapidly to the long-wavelength side, whereas the
1\textit{s}-exciton absorption line stays almost constant, due to a
high degree of compensation between the weakening of the
electron-hole binding energy and the band-gap reduction, like that
earlier found for GaAs \cite{{Haug}}. At high electron-hole
concentrations, the electric dipole moment renormalization effects
give rise to a large optical gain which is shown in Figs. 5 and 6.
The maximum of the Hartree--Fock gain spectrum equal to $29991.2~
\textrm{cm}^{-1}$ is observed at the wavelength  $\lambda=339.1$ nm
at a concentration of $9\times 10^{12}~\textrm{cm}^{-2}$. From Fig.
6, we see that the Hartree--Fock spectrum is shifted to the
long-wavelength side relative to the Hartree gain spectrum.
Moreover, a large Sommelfeld or Coulomb enhancement is present in
the quantum well. It is caused by an increase of the oscillator
strength due to the electron-hole Coulomb attraction.

The exchange Hartree--Fock energy spectrum
renormalization is accounted in the equation of motion for the microscopic
dipole of the electron-hole pair. For high concentrations, this value
is significant. It is somewhat larger for electrons and less for
holes. Totally, this value is reflected in the Hartree--Fock gain
shifting in comparison with the Hartree spectrum, as shown in Fig. 6. It
should be noted that the gain spectrum involves not only
Hartree--Fock correlations, but correlations of higher orders
in the expansion in the Coulomb potential energy. This
is achieved by the summation of the series in the Coulomb energy in the microscopic polarization
equation in all orders of perturbation theory. In more
details, the microscopic polarization equation for the dipole of an
electron-hole pair for the w\"{u}rtzite quantum well will be considered in
our next paper.

\begin{figure}
\includegraphics[width=\column]{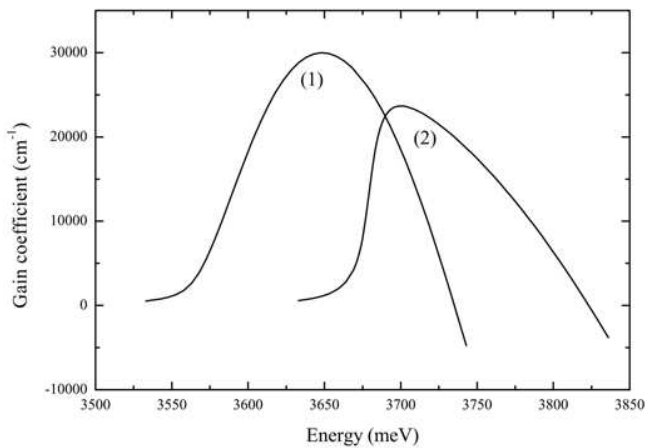}
\vskip-3mm\caption{Calculated Hartree--Fock gain spectrum for the
quantum well 2.6 nm in width at the concentration $n=p=9\times
10^{12}~\textrm{cm}^{-2}$ (1) and the Hartree spectrum at the
concentration $n=p=9\times 10^{12}~\textrm{cm}^{-2}$ (2)}
\end{figure}

\section{Conclusions}

In summary, the self-consistant calculations of the Schr\"{o}dinger
equations and the Poisson equation of wide bandgap GaN/AlGaN quantum
well systems show the interesting dependences of the matrix elements for dipole optical
interband transitions and the light gain spectrum on the quantum well width and the charge
density. A blueshift with
increase in the plasma density in the gain spectrum in relatively wide wells occurs as
a consequence of the screening of the piezoelectric field induced
by the quantum confined Stark effect, whereas the structures with narrow well
widths exhibit the usual dependence of the development of the light gain
maximum almost without shifting the spectral region. It is found that
the matrix elements of optical transitions from the heavy hole band
have the strict TE light
polarization like that of the light gain spectrum. With regard for the Coulomb interactioin, a red
shift of the Hartree--Fock light gain spectrum relative to the Hartree gain spectrum and a large
Sommerfeld enhancement for a quantum well are found.

\vskip3mm The author is grateful to Prof. V.I.~Sheka and Prof.
V.A.~Kochelap for numerous discussions.

\vskip4mm
\rezume{%
ОПТИЧНА~~ ПОЛЯРИЗАЦІЙНА~~ АНІЗОТРОПІЯ,\\ВНУТРІШНІЙ ЕФЕКТ ШТАРКА
КВАНТОВОГО\\ КОНФАЙНМЕНТУ І ВПЛИВ КУЛОНІВСЬКИХ \\ЕФЕКТІВ НА ЛАЗЕРНІ
ХАРАКТЕРИСТИКИ\\
$[0001]$-ОРІЄНТОВАНИХ GaN/Al$_{0,3}$Ga$_{0,7}$N\\
КВАНТОВИХ ЯМ }{Л.O. Локоть} {У цій статті представлено теоретичне
дослідження про\-сто\-ро\-во розділених електронних і діркових
розподілів, яке ві\-доб\-ра\-жа\-єть\-ся у самоузгодженому
розв'язанні рівнянь Шредінгера для електронів та дірок і рівняння
Пуассона. Результати проілюст\-ро\-ва\-но для
GaN/Al$_{0,3}$Ga$_{0,7}$N квантової ями. Спектр оптичного
підси\-ле\-ння в [0001]-орієнтованої GaN/Al$_{0,3}$Ga$_{0,7}$N
квантової ями  обчислено в ультрафіолетовій області. Знайдено, що як
мат\-рич\-ні елементи оптичних переходів з важкої діркової підзони в
зону провідності, так і спектр оптичного підсилення мають строго
\textit{x }(або \textit{y}) поляризацію світла. Показано вплив
конфайнменту хвильових функцій на оптичне під\-си\-ле\-ння, яке
неявно залежить від вбудованого електричного поля, що обчислене і
дорівнює 2,3 MВ/cм. Якщо структури з вузькою шириною ями проявляють
звичайну залежність роз\-вит\-ку максимуму під\-си\-ле\-ння світла
майже без зміщення спектра\-льної області, то значного голубого
зміщення максимуму під\-си\-ле\-ння зі зростанням густини плазми
набувають структури зі значною шириною квантової ями. Це голубе
зміщення від\-но\-сять до взаємодії між екрануючим п'єзоелектричним
полем, створеним деформацією і зонною структурою. Велике
зоммер\-фель\-дів\-ське або кулонівське підсилення присутнє у
квантовій ямі.}

\end{document}